\documentclass{article}
\usepackage{spconf,amsmath,epsfig}
\usepackage{pifont}

\usepackage{subfig}
\usepackage{booktabs}
\usepackage{array}
\usepackage{float}
\usepackage{amssymb}
\usepackage{color}
\usepackage{multirow,cite}
\usepackage[multiple]{footmisc}

\let\OLDthebibliography\thebibliography
\renewcommand\thebibliography[1]{
  \OLDthebibliography{#1}
  \setlength{\parskip}{0pt}
  \setlength{\itemsep}{0pt plus 0.3ex}
}

\begin{document}\sloppy

\def\x{{\mathbf x}}
\def\L{{\cal L}}

\title{AudioLog: LLMs-Powered Long Audio Logging with \\Hybrid Token-Semantic Contrastive Learning}
%


\name{
Jisheng Bai{\rm\textsuperscript{1,3}}, Han Yin{\rm\textsuperscript{1}}, Mou Wang{\rm\textsuperscript{2}}, Dongyuan Shi{\rm\textsuperscript{3}}, Woon-Seng Gan{\rm\textsuperscript{3}}, Jianfeng Chen{\rm\textsuperscript{1}}, Susanto Rahardja{\rm\textsuperscript{1}}
}

\address{
	\textsuperscript{1} School of Marine Science and Technology, Northwestern Polytechnical University, Xi'an, China\\
    \textsuperscript{2} Institute of Acoustics, Chinese Academy of Sciences, Beijing, China\\
    \textsuperscript{3} School of Electrical \& Electronic Engineering, Nanyang Technological University, Singapore.
}

\maketitle

\begin{abstract}
Previous studies in automated audio captioning have faced difficulties in accurately capturing the complete temporal details of acoustic scenes and events within long audio sequences.
This paper presents AudioLog, a large language models (LLMs)-powered audio logging system with hybrid token-semantic contrastive learning. 
Specifically, we propose to fine-tune the pre-trained hierarchical token-semantic audio Transformer by incorporating contrastive learning between hybrid acoustic representations. 
We then leverage LLMs to generate audio logs that summarize textual descriptions of the acoustic environment.
Finally, we evaluate the AudioLog system on two datasets with both scene and event annotations.
Experiments show that the proposed system achieves exceptional performance in acoustic scene classification and sound event detection, surpassing existing methods in the field.
Further analysis of the prompts to LLMs demonstrates that AudioLog can effectively summarize long audio sequences\footnote{Code is released in: https://github.com/JishengBai/AudioLog}.
To the best of our knowledge, this approach is the first attempt to leverage LLMs for summarizing long audio sequences. 
\end{abstract}
\section{Introduction}
\label{sec:intro}

Audio pattern recognition (APR) is an expanding field of signal processing and machine learning techniques to comprehend the environment that surrounds us.
Many audio analysis tasks have emerged and been deeply investigated, such as acoustic scene classification (ASC) \cite{virtanen2018computational}, audio tagging (AT) \cite{bai2022multimodal}, sound event detection (SED) \cite{mesaros2021sound}, automated audio captioning (AAC) \cite{mei2023wavcaps}, etc.
Among the tasks, AAC, a relatively new concept, has recently garnered significant attention \cite{xu2022comprehensive}.
It introduces the novel idea of generating textual descriptions for audio recordings by combining audio signal processing and natural language processing techniques.
However, previous studies have primarily focused on developing captioning for short audio clips, the captioning capabilities for longer audio sequences have not been studied. 

Long audio sequences, abundant in acoustic scene and event data, offer a wealth of information for environmental sensing.
Summarizing long audio sequences can be potentially used in safety monitoring for children, elderly and hearing-impaired people, and
improving the efficiency of editing for content creators.
Yet, dealing with such an amount of information has been challenging for AAC systems in the past.
In recent years, large language models (LLMs) like GPT-4, ChatGPT, and Llama have shown amazing abilities in handling a wide range of tasks \cite{touvron2023llama, brown2020language}.
Therefore, we can leverage the powerful abilities of LLMs to process and summarize acoustic information for long audio signals.

Previous AAC systems have incorporated large pre-trained audio models to improve performance, but the systems do not comprehensively investigate the temporal information modeling of acoustic scenes and events \cite{mei2022automated, xu2022comprehensive, kong2020panns, chen2022hts}. 
Multi-task learning (MTL) has been integrated into the training of audio models to enable the temporal prediction of environmental sounds \cite{imoto2020sound, bai2022squeeze}.
Nonetheless, modeling the temporal relationships of acoustic scenes and events simultaneously is still challenging.
Recently, contrastive learning has been used in several audio tasks to learn effective and general acoustic representations \cite{al2021clar,wang2021multi,wu2023large}.
Therefore, we can leverage contrastive learning to extract more comprehensive representations, and further improve the performance for predicting the temporal information of acoustic scenes and events.

In this paper, we introduce AudioLog, a state-of-the-art logging system powered by LLMs for summarizing long audio sequences. 
Firstly, we propose a hybrid token-semantic contrastive learning framework to fine-tune the pre-trained hierarchical token-semantic audio Transformer (HTS-AT).
Specifically, the HTS-AT is fine-tuned using an MTL approach that incorporates ASC, SED, and contractive learning of hybrid acoustic representations.
Subsequently, the fine-tuned model is used to produce temporal information of acoustic scenes and events simultaneously.
Finally, we leverage LLMs to summarize textual descriptions of the acoustic information for long audio sequences by prompting LLMs with different inquiries.
Experimental results show that the AudioLog system outperforms the compared methods for ASC and SED. 
Further analysis regarding the generation of audio logs indicates that AudioLog can effectively summarize the acoustic contents for long audio sequences.

\begin{figure*}[t!]
	\centering
	\includegraphics[width=16cm]{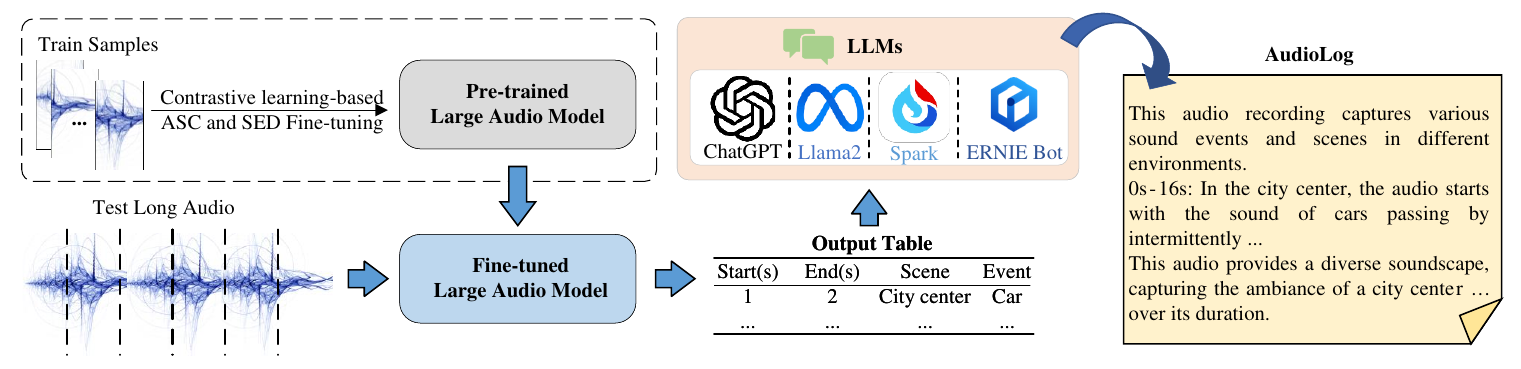}
	\caption{The overview of the proposed AudioLog system.}
	\label{fig:audiolog overview}
\end{figure*}

\section{Proposed method}
\label{sec:method}

\subsection{Overview}

The processing pipeline of the proposed AudioLog system is illustrated in Fig. \ref{fig:audiolog overview}.	
For training, the audio segments are used for fine-tuning the large audio model with hybrid contrastive learning on ASC and SED.
For testing, the long test audio is segmented and processed by the fine-tuned large audio model to generate text contents of acoustic scenes and events with temporal information.
The text contents are then organized into a table with columns of "Start", "End", "Scene", and "Event". 
Here, "Start" and "End" denote the respective start and end times (in seconds) for the corresponding scenes and events.
Finally, the organized table is fed into the LLMs, which are prompted to summarize the audio by incorporating temporal information and providing a concise description, referred to as AudioLog.

\begin{figure}[t!]
	\centering
	\includegraphics[width=6cm]{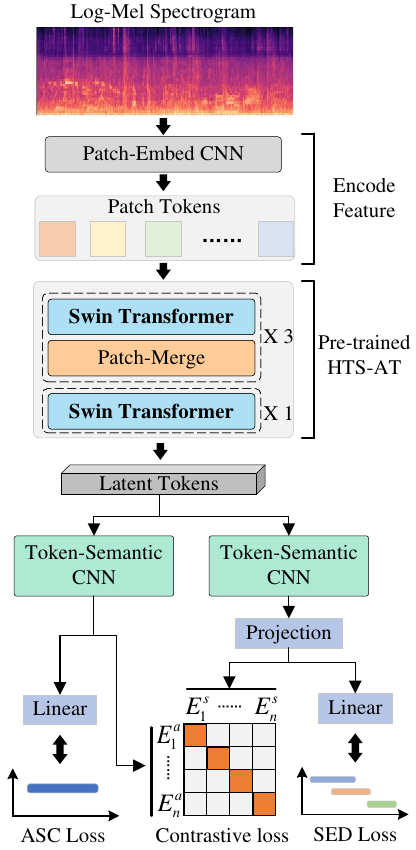}
	\caption{The architecture of hybrid token-semantic contrastive learning framework for fine-tuning the pre-trained hierarchical token-semantic audio Transformer (HTS-AT).}
	\label{fig:MTL}
\end{figure}

\subsection{Hybrid token-semantic contrastive learning}

The proposed hybrid token-semantic contrastive learning framework based on the pre-trained HTS-AT is shown in Fig. \ref{fig:MTL}.
This network first encodes the acoustic features into patch tokens, which are subsequently fed into the pre-trained HTS-AT.
Then HTS-AT outputs latent tokens, which are fed into ASC and SED token-semantic CNN branches.
Specifically, we adopt the token-semantic CNNs for generating two hybrid embeddings, which represent the coarse and fine acoustic representations for the ASC and SED branches, respectively, 
The hybrid embeddings are used to calculate the contrastive loss to enhance the learning of general acoustic representations.
Finally, we train the model by combining the ASC, SED and contrastive losses.

For feature encoding, the input audio is first transformed into a log-Mel spectrogram of $T$ frames and $F$ bins, which is then segmented into $(\frac{T}{P} \times \frac{F}{P},\, D)$ patch tokens using a patch-embed CNN with a kernel of $(P \times P)$, where $D$ is the latent state dimension.
Then the patch tokens are fed into 4 network groups, achieving latent tokens with the shape of $(\frac{T}{8P} \times \frac{F}{8P},\, 8D)$.
Each of the first three groups consists of a Swin Transformer \cite{liu2021swin} block with a shifted window attention and a patch-merge layer, while the fourth only contains a Swin Transformer block.

After that, we introduce ASC and SED token-semantic CNN-based branches.  
The ASC token-semantic CNN has a channel number of 128, a kernel size of $(1, \frac{T}{8P})$, and a padding size of $(0, 0)$, outputting the embedding $E^{a}\in \mathbb{R}^{128}$.
The SED token-semantic CNN has the same channel number as ASC but with a kernel size of $(3, 3)$, and a padding size of $(1\, 1)$, outputting features of $(128, \frac{T}{8P})$.
We then use a linear layer as the projection layer to get the embedding $E^{s}\in \mathbb{R}^{128}$.

Finally, we use a linear layer to get the output vector for calculating the cross-entropy loss of ASC, and a linear layer to get the event presence map for calculating the binary cross-entropy loss of SED.
The embeddings are used to calculate contrastive loss $\mathcal{L}_{c}$ between the ASC and SED token-semantic CNN-based branches, expressed as:
\begin{equation}
\begin{split}
\mathcal{L}_{c} =& \frac {1}{2N} \sum_{i=1}^ {N} (\log \frac {exp(E_{i}^{a}\cdot E_{i}^{sT}/\tau )} {\sum_{j=1}^{N}exp(E_{i}^{a}\cdot E_{j}^{s\mathrm{T}}/\tau )} \\ &+ 
\log \frac {exp(E_{i}^{s}\cdot E_{i}^{aT}/\tau )} {\sum_{j=1}^{N}exp(E_{i}^{s}\cdot E_{j}^{a\mathrm{T}}/\tau )})
\end{split}
\end{equation}
where $E_{i}^{a}$ and $E_{i}^{s}$ are embedding from ASC and SED branch for the same input index $i$, $N$ is the batch size, and $\tau$ is a learnable temperature parameter for scaling the loss.
The model is trained with the ASC, SED, and contrastive losses:
\begin{equation}
\mathcal{L} = \alpha\mathcal{L}_{a}+\beta \mathcal{L}_{s}+(1-\alpha-\beta)\mathcal{L}_{c}
\end{equation}
where $\alpha$ and $\beta$ are weight factors, $\mathcal{L}_{a}$ and $\mathcal{L}_{s}$ denote the losses of ASC and SED, respectively.

\subsection{LLMs-based AudioLog}
To predict long test audio sequences, we establish a unified output format that includes both SED and ASC results, represented as $(S, E, C_{e}, C_{s})$. 
In this context, $S$ and $E$ represent the respective start and end times of the categories $C_{s}$ (scenes) and $C_{e}$ (events).
The test audio is segmented to serve as input for fine-tuning the large audio model. 
The model outputs are then concatenated into an output table, which records the start and end times, as well as the events and scenes occurring throughout the audio.
Subsequently, LLMs utilize the output table as input, following the prompts to produce the ultimate AudioLog. 
Table \ref{tab:example table} illustrates an example of the structure of the output table.

\begin{table}[h]
\centering
\caption{An example of the output table for joint estimating ASC and SED results.}
\label{tab:example table}
\begin{tabular}{cccc}
\toprule
Start(s) & End(s) & Scene  & Event          \\ \hline
0 & 1  & city\_center      & car            \\
... & ...    & ...  & ...            \\
16  & 17     & metro\_station    & metro leaving  \\
... & ...    & ...               & ...            \\
40  & 41     & residential\_area & birds\_singing \\
... & ...    & ...               & ...            \\
59 & 60     & residential\_area & birds\_singing \\ 
\bottomrule
\end{tabular}
\end{table}

\section{Experiments}
\label{sec:Experiments}
\subsection{Dataset}
We use two datasets with both scene and event labels to evaluate the proposed AudioLog.
The first dataset is the development dataset of DCASE 2023 Task 4B. 
It comprises real-life audio recordings, each approximately 3 minutes in length, captured across 5 acoustic scenes and 11 event classes \cite{Martinmorato2023, Martinmorato2023b}.
The second dataset is derived from the joint analysis of sound events and acoustic scenes\cite{igarashi2022information, imoto2020sound}. 
This dataset consists of segments extracted from the TUT Acoustic Scenes 2016\&2017 and TUT Sound Events 2016\&2017 datasets. 
The audio clips are annotated with 4 acoustic scenes and 25 sound events.
Further details regarding these datasets can be found in the websites\footnote{https://dcase.community/challenge2023/task-sound-event-detection-with-soft-labels}\footnote{https://www.ksuke.net/dataset/strong-sound-event-labels-of-tut-acoustic-scenes-2016-2017}.

\subsection{Settings}
We followed the experimental configurations of HTS-AT while fine-tuning the model. 
We used the short-time Fourier transform with a window size of 1024 and a hop length of 320 to generate spectrograms from the audio signals.
Mel filters of 64 bands are used to calculate Mel spectrograms. 
The network architecture comprises 4 distinct groups, featuring 2, 2, 6, and 2 Swin-Transformer blocks, respectively.
We use the AdamW optimizer with a learning rate of 0.0001 and a batch size of 32.
$\tau$ is set to 2.6592.
The weight factors $\alpha$ and $\beta$ for combining the losses are set to $0.3$ and $0.6$, respectively. 
The evaluation metrics are the average of the class-wise accuracies (ACC) for ASC, error rate (ER), and segment-based micro-average F1 score (F1\_m) for SED.
The last update date for all the LLMs tested in this article is September 14, 2023.

\subsection{Results}


\subsubsection{Overall performance of ASC and SED}

\begin{table}[h]
\centering
\renewcommand\arraystretch{1.1}	
\caption{ASC and SED Performance Comparison on Different Datasets and Methods. CL denotes contrastive learning.}
\label{tab:Results of ASC and SED}
\begin{tabular}{>{\centering\arraybackslash}p{1.7cm}>{\centering\arraybackslash}p{2.7cm}>{\centering\arraybackslash}p{0.7cm}>{\centering\arraybackslash}p{0.7cm}>{\centering\arraybackslash}p{0.7cm}}
\toprule
\multirow{2}{*}{Dataset} & \multirow{2}{*}{Method} & ASC   & \multicolumn{2}{c}{SED} \\
                         &            & ACC$\uparrow$  & ER$\downarrow$   & F1\_m$\uparrow$         \\ \hline
         & Baseline \cite{Martinmorato2023b}       & -     & 0.487      & 0.703      \\
DCASE2023 & Top-2 \cite{Jin2023}	     & -	 & 0.430 	  & 0.729      \\
Task 4B   & Top-1 \cite{Yin2023}   & -     & 0.360      & 0.786      \\
    & Proposed (w/o CL)  & 0.884 & \textbf{0.300} & 0.807      \\    
    & Proposed (w/ CL)   & \textbf{0.897} & \textbf{0.300} & \textbf{0.836}      \\
    \hline
\multirow{4}{*}{TUT16\&17} & CNN \cite{nada2023joint}        & 0.843 & -          & -          \\
                         & CNN-BiGRU\cite{nada2023joint}  & -     & -          & 0.425      \\
                         & Imoto et al.\cite{nada2023joint} & 0.894 & -          & 0.461      \\
    & Proposed (w/o CL) & 0.942 & 0.767  & 0.495     \\ 
    & Proposed (w/ CL) & \textbf{0.947} & \textbf{0.727} & \textbf{0.505}      \\
\bottomrule
\end{tabular}
\end{table}

Table \ref{tab:Results of ASC and SED} shows the ASC and SED results of the proposed methods.
For the DCASE 2023 Task 4B dataset, we compared three methods, including the baseline, Top-2's, and Top-1's methods of the challenge \cite{Martinmorato2023b, Jin2023, Yin2023}.
The proposed methods outperform all other methods on this dataset.
Moreover, the introduction of contrastive learning in the proposed method further improves the performance, achieving an ACC of 0.897, an ER of 0.300, and an F1\_m score of 0.836. 
For the TUT16\&17 dataset, we compared three methods: CNN, CNN-BiGRU, and Imoto's method \cite{nada2023joint}.
Our methods surpass all other methods, and the contrastive learning-based model achieves the best performance with an ACC of 0.947, an ER of 0.727, and an F1\_m score of 0.505.
The experimental results indicate that the proposed methods consistently outperform existing state-of-the-art approaches on both datasets.
In addition, the proposed hybrid token-semantic contrastive learning framework further improves the performance of joint estimation on ASC and SED.

\subsubsection{Results of different prompts}
\begin{table}[h!]
\centering
\caption{The answers of ChatGPT for various prompts in summarizing contents of acoustic events and scenes.}
\label{tab:results of ChatGPT}
\begin{tabular}{cccc}
\toprule
\multicolumn{1}{p{1cm}}{\textbf{General prompt}} & \multicolumn{3}{p{6cm}}{The above table provides a description of acoustic events and scenes from an audio clip, along with their start and end times in seconds. }  \\ \hline
\multicolumn{1}{p{1cm}}{\textbf{Prompt1}} & \multicolumn{3}{p{6cm}}{Please provide a concise overview of this audio.} \\
\\ 
\multirow{5}{*}{\textbf{Answer1}} & \multicolumn{3}{p{6cm}}{The provided audio clip consists of several scenes and events, each with their respective start and end times in seconds. Here is a summary of the audio content:} \\
& \multicolumn{3}{p{6cm}}{\textbf{City Center Scene:}} \\
& \multicolumn{3}{p{6cm}}{Car sounds from... 

Large vehicle sounds...}\\
& \multicolumn{3}{p{6cm}}{\textbf{Metro Station Scene: ...}}\\                  
& \multicolumn{3}{p{6cm}}{In summary, the audio clip transitions between scenes in a city center (with cars...} \\ 
\hline
\multicolumn{1}{p{1cm}}{\textbf{Prompt2}} & \multicolumn{3}{p{6cm}}{Please provide a concise overview of this audio, along with the timing information for sound scenes and events.}                        \\ \\
\multirow{5}{*}{\textbf{Answer2}} & \multicolumn{3}{p{6cm}}{This audio clip can be summarized as follows, with relevant time descriptions:} \\
& \multicolumn{3}{p{6cm}}{\textbf{City Center Scene} (0-16 seconds):} \\
& \multicolumn{3}{p{6cm}}{Various car sounds can be heard intermittently from...}\\
& \multicolumn{3}{p{6cm}}{\textbf{Metro Station Scene} (16-36 seconds): ...}\\  
& \multicolumn{3}{p{6cm}}{This summary provides an overview of the different scenes and events along with their respective time descriptions.} \\ 
\hline
\multicolumn{1}{p{1cm}}{\textbf{Prompt3}} & \multicolumn{3}{p{6cm}}{Please provide a concise overview of this audio without timing information.}                    \\    \\ 
\multirow{5}{*}{\textbf{Answer3}} & \multicolumn{3}{p{6cm}}{This audio clip is divided into different scenes and events. Here's a summary of the audio without the time descriptions:} \\
& \multicolumn{3}{p{6cm}}{In the "city\_center" scene:} \\
& \multicolumn{3}{p{6cm}}{There are alternating sounds of "car" ...}\\
& \multicolumn{3}{p{6cm}}{In the "metro\_station" scene: ...}\\ 
& \multicolumn{3}{p{6cm}}{Overall, the audio clip depicts various urban and suburban scenes with a mix of transportation sounds, human activity, and natural elements like birds singing.} \\ 
\hline
\end{tabular}
\end{table}

To explore the influence of prompts at different time scales, we conducted experiments on one of the LLMs, ChatGPT, and the results are shown in Table \ref{tab:results of ChatGPT}.
We used a predicted table of acoustic scenes and events from an urban audio recording with a length of approximately 3 minutes as the input to ChatGPT.

Initially, we provided a general prompt describing the structured output table of acoustic temporal information. 
Then, we gave Prompt1, which requested only a concise overview.
In response to Prompt1, ChatGPT offers a comprehensive overview of the audio content from acoustic scene aspects, along with the corresponding start and end times of sound events. 
For prompt2, we added a requirement to provide timing information
for sound scenes and events. 
Similar to Answer1, ChatGPT describes the corresponding start and end times of sound events but also describes the temporal information of the acoustic scenes.
For Prompt3, we focus on a more concise summary of the audio without specifying timing details.
Answer3 highlights the key sound events within each scene, offering a broader overview of the audio while omitting specific time-related details.
The answers of different prompts illustrate that LLMs can provide a flexible and comprehensive description of long audio sequences based on the specific information needs of the user.

\subsubsection{Audio logs of different LLMs}

\begin{figure}[t]
	\centering
	\includegraphics[width=8.5cm]{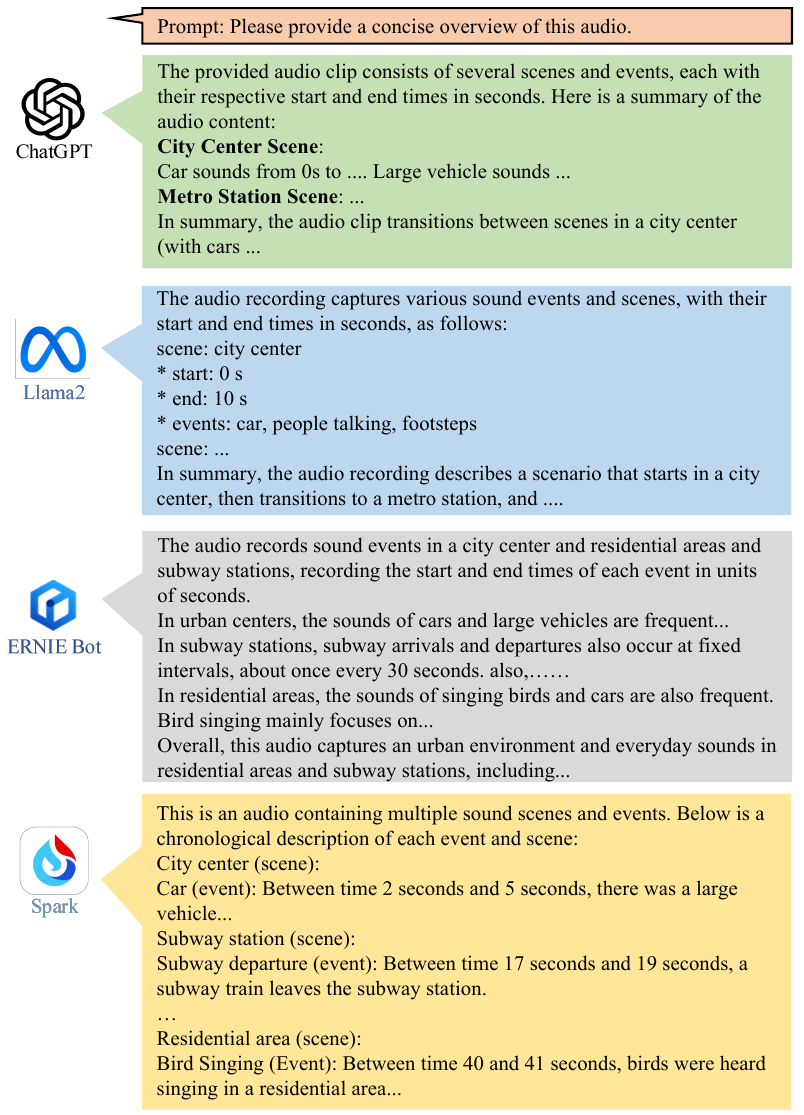}
	\caption{Examples of audio logs for different LLMs. The logs are generated from a predicted table of acoustic scenes and events from an urban audio recording with a length of approximately 3 minutes.}
	\label{fig:audiolog example}
\end{figure}

To investigate the summarizing capabilities of different LLMs for acoustic information in long audio sequences, we conducted experiments using several popular LLMs, including ChatGPT, Llama2, Spark, and ERNIE Bot.
The characteristics of each LLM in summarizing the environmental sound contexts are illustrated in the following:

ChatGPT: ChatGPT offers a summary of the audio content with acoustic scenes and respective start and end times of events in seconds. 

Llama2: Llama2 provides a structured description of the audio, detailing the start and end times of scenes and associated sound events. 

ERINE Bot: ERINE Bot outlines the acoustic scenes and provides the temporal information of sound events in seconds. It highlights some events in different scenes.

Spark: Spark offers a time-based description of sound events item by item and organizes them with different scenes. 

The above results demonstrate that while the LLMs have differences in training methods and styles, all of the LLMs are effective in summarizing the acoustic contents for long audio sequences.
Based on the experimental results, the authors recommend ChatGPT because it provides a comprehensive overview of long audio content while effectively describing the details of sound events.

\section{Conclusion}
\label{sec:conclusion}

This paper introduced AudioLog, an LLMs-powered logging system for long audio sequences. 
We proposed a hybrid token-semantic contrastive learning framework to fine-tune the pre-trained HTS-AT.
This framework can enhance the learning of general acoustic representations and further improve the performance of extracting temporal information from scenes and events in the acoustic environment. 
Moreover, we explored the recent popular LLMs in summarizing the textual information by prompting in different scales.
Experimental results show that the proposed AudioLog system can effectively summarize the temporal information related to acoustic scenes and events in long audio sequences.

\bibliographystyle{IEEEbib}
\bibliography{icme2023template}

\end{document}